# Design Omnidirectional Wave Absorbers by Transformation Method

**Zheng Chang** and **Gengkai Hu**[*]

[1]*Key Laboratory of Dynamics and Control of Flight Vehicle, Ministry of Education, School of Aerospace Engineering, Beijing Institute of Technology, 100081,Beijing, China.*

**Abstract**: A general conformal mapping is proposed to design omnidirectional broadband wave absorbers by transformation method. When applied to electromagnetic (EM) and acoustic waves, the existing material parameters of the EM and acoustic omnidirectional absorbers, which are previously obtained by Hamiltonian optics and geometry acoustics, can be recovered. In addition, magnetic and mass-density-controlled omnidirectional absorbers for EM and acoustic waves can also be designed, respectively. We then apply the conformal mapping to design an omnidirectional elastic wave absorber, the corresponding material realization of such elastic absorber is also proposed and validated by numerical simulation.

[*]Corresponding author, hugeng@bit.edu.cn

Energy absorption and collection are important issues in various engineering applications. Usually, energy flows should be directed into a region where it is absorbed or taken away. An omnidirectional absorber, also called "optical black-hole", is recently designed based on Hamiltonian optics for such purpose [1], and the concept is then validated experimentally through different techniques [2-4]. By analogy, similar acoustic devices are designed both analytically [5] and experimentally [6]. However, this analogy does not hold for elastic waves in solids, therefore a more general method is necessary in order to design elastic wave absorber. The recently developed transformation method can systematically establish the relation between material distribution and function of a device for different waves, such as electromagnetic [7-9], acoustic [10-12], matter waves [13] and even surface water waves [14]. It is also extended approximately to elastic waves [15] and fluids dynamics [16]. It is natural to imagine if we can find an appropriate mapping representing correctly the function of the "black-hole" device, omnidirectional absorbers corresponding to the above-mentioned different waves can then be designed directly. This is objective of the work.

In this letter, a general design approach for "black-hole" absorption devices will be proposed based on transformation method. With help of a series of conformal mappings, the geometry space of such device is determined, and the necessary material distribution of the device is obtained with the transformation relation specific to different physical processes. It is shown that the existing "black-hole" devices [1, 5] can be recovered from the proposed transformation approach. In addition, a magnetic "optical black-hole" and a mass-density-controlled "acoustic black-hole" can also be proposed. Finally, an elastic wave "black-hole" device is proposed, and the prototype of such device is also designed and validated by numerical simulation.

We start by considering the following conformal mapping

$$w = Az^n \ (A \neq 0, n < 0), \tag{1}$$

in which $z = x + iy$ represents the virtual space and $w = x' + iy'$ represents the physical space. When $n \in (-\infty, 0)$ (See Fig.1 as an example), a flat virtual space is mapped into a curved one, and at the same time, the infinity of the virtual space is mapped into the origin of the physical space, this fact reveals its potential to be used in designing "black-hole" devices.

The conformal mapping keeps the isotropic property of the transformed material if the material in the virtual space is also isotropic [8, 17, 18]. In this context, the Jacobian determinant $J$ can be used as the only variable connecting the space and material. Specific for the mapping (1), the Jacobian determinant is obtained as:

$$J = An(\frac{1}{A}r')^{\frac{2(n-1)}{n}}, \tag{2}$$

where $r' = \sqrt{x'^2 + y'^2}$ is the polar coordinate of the physical space.

The outer boundary of the device is required to be matched with the surrounding space, it is known that in the place where $J = 1$ from the mapping (1), there is no transformation between the virtual space and physical space. Therefore, we use $J = 1$ as a constraint condition to determine the outer boundary of the device. After simple calculation, a circular geometry of the device with a radius $R'$ is found as

$$R' = \left| A(An)^{\frac{n}{2(1-n)}} \right|. \tag{3}$$

Based on this result, inserting Eq. (3) into Eq. (2), a more concise variation of the Jacobian determinant as function of $R'$ and $n$ is obtained as:

$$J = (\frac{R'}{r'})^{-2+\frac{2}{n}}, n \in (-\infty, 0). \tag{4}$$

With Eq. (4), a series of "black-hole" devices can be designed with help of transformation relations for different physical processes. For example, in "transformation optics", the transformation relations for 2-D TM problems are given by

$$\boldsymbol{\mu}' = \frac{\mathbf{AA}^T}{J}\mu, \varepsilon' = \frac{1}{J}\varepsilon, \quad (5)$$

where $\mathbf{A}$ is the Jacobin matrix of the mapping. A non-magnetic "optical black hole" can be designed by inserting the Cauchy-Riemann condition and Eq. (4) into Eq. (5). Finally, the material parameters of the non-magnetic "optical black hole" has the following form

$$\mu' = \mu, \varepsilon' = (\frac{R'}{r'})^{2-\frac{2}{n}}\varepsilon, (r' \leq R'). \quad (6)$$

Similarly, in "transformation acoustics", with the transformation relation reported in [10]:

$$\boldsymbol{\rho}^{-1}{}' = \frac{\mathbf{AA}^T}{J}\rho^{-1}, \kappa' = J\kappa, \quad (7)$$

"acoustic black hole" with the following material parameters

$$\rho' = \rho, \kappa' = (\frac{R'}{r'})^{-2+\frac{2}{n}}\kappa, (r' \leq R'), \quad (8)$$

can be designed, they are the same as those reported before [1, 5] obtained from a different approach. In addition, a magnetic "optical black-hole" can also be designed by considering the transformation relation of TE problems in "transformation optics":

$$\boldsymbol{\varepsilon}' = \frac{\mathbf{AA}^T}{J}\varepsilon, \mu' = \frac{1}{J}\mu. \quad (9)$$

The material distribution of the magnetic "optical black-hole" reads

$$\varepsilon' = \varepsilon, \mu' = (\frac{R'}{r'})^{2-\frac{2}{n}}\mu, (r' \leq R'). \quad (10)$$

With the transformation relation for acoustic wave reported in [19], an "acoustic black-hole" device with one constant material parameter can be designed. If the bulk modulus is assumed to be constant, we can choose the following transformation relations [19]:

$$\boldsymbol{\rho}^{-1}{}' = \mathbf{A}\mathbf{A}^T \rho^{-1}, \kappa' = \kappa, \tag{11}$$

which lead to the following mass-density-controlled "acoustic black-hole"

$$\rho' = (\frac{R'}{r'})^{2-\frac{2}{n}} \rho, \kappa' = \kappa, (r' \leq R'), \tag{12}$$

To avoid singular point and large gradient of material parameters, we can replace the center part of the device $(r' \leq R_c')$ with an homogeneous material [1-6],

$$\varepsilon' = \varepsilon + i\delta, \mu' = (\frac{R'}{R_c'})^{2-\frac{2}{n}} \mu, (r' \leq R_c'), \tag{13}$$

for the magnetic "optical black-hole" and

$$\rho' = (\frac{R'}{R_c'})^{2-\frac{2}{n}} \rho, \kappa' = \kappa + i\delta, (r' \leq R_c') \tag{14}$$

for the mass-density-controlled "acoustic black-hole", where $\delta$ is loss coefficient. FEM simulations for these two devices are carried out. The geometry parameters of the devices are set to be $R' = 12\text{mm}$, $R_c' = 4\text{mm}$, $\delta = 0.6$ and $n = -\infty$ in order to facilitate the comparison with the previous works [1, 5]. It's shown in Fig. 2(a) that the Gaussian beam with a frequency $f = 100\text{GHz}$ impinging on the magnetic "black-hole" is bended into the shell region and then absorbed by the core, the similar phenomenon is also observed for the designed mass-density-controlled "acoustic black-hole", as shown in Fig. 2(b). Other transformation relations can also be proposed to design different kinds of "acoustic black-holes", more details can be found in [12, 19, 20].

To further illustrate the validity of the above method, an elastic wave "black-hole" device will be designed. Therefore in the following, we will apply the proposed mapping to elastodynamics problems [15, 18]. As been reported in [18], there are different forms of the transformation relations in transformation elastodynamics for a conformal mapping. Here we choose arbitrarily the following relations

$$\rho' = J^{-\frac{1}{2}}\rho,\ E' = J^{\frac{1}{2}}E,\ \upsilon' = \upsilon \tag{15}$$

where $E$ is the Young's modulus and $\upsilon$ is the Poisson's ratio. Inserting Eq. (4) into (15), the material distributions of the elastic wave "black-hole" are derived:

$$\rho' = (\frac{R'}{r'})^{1-\frac{1}{n}}\rho,\ E' = (\frac{R'}{r'})^{-1+\frac{1}{n}}E,\ \upsilon' = \upsilon,\ (r' \leq R'). \tag{16}$$

Also, the core region $(r' \leq R_c')$ is set to be

$$\rho' = (\frac{R'}{R_c'})^{1-\frac{1}{n}}\rho,\ E' = (\frac{R'}{R_c'})^{-1+\frac{1}{n}}E(1+i\delta),\ \upsilon' = \upsilon,\ (r' \leq R_c'). \tag{17}$$

As the required, the material parameters of the elastic "black-hole" device are isotropic, the implementation turns to be very simple. Here, a detailed scheme to construct the above elastic wave "black-hole" will be examined. In the case where the background material is an aluminum foam ($\rho = 0.4\ \text{kg/m}^3$, $E = 2\ \text{GPa}$, $\upsilon = 0.27$ and $\delta = 0.003$ [21]), and the parameters of the "black-hole" device are $R' = 1\text{m}$, $R_c' = 0.5\text{m}$, $n = -1$ and $\delta = 0.6$, the "black-hole" region can be discretized into 10 layers with the same thickness. Each layer is spliced by the cell element as shown in Fig. 3(a). In the quasi-static approximation, the effective material parameters follow the mixing rule, therefore the required material parameters in each layer can be fulfilled by adjusting the radius of the circular inclusion of each cell element. As the matrix is aluminum foam, we chose silicone ($\rho = 2.3\ \text{kg/m}^3$, $E = 0.05\ \text{GPa}$, $\upsilon = 0.47$ and $\delta = 1.6$ [21]) and melamine foam ($\rho = 0.012$ kg/m$^3$, $E = 0.0004$ Gpa, $\upsilon = 0.2$ and $\delta = 0.5$ [21]) as the materials for the inclusions, in order to adjust two material parameters ($\rho'$ and $E'$) simultaneously. The Poisson's ratio $\upsilon$ is not controlled in this model. FEM simulations are carried out on this discrete model to illustrate the effectiveness of the device. To reduce the computation cost, the homogeneous core of the "black-hole" device is set to be an ideal material followed from Eq. (17). As shown in figure 3(b), an S-wave beam with $f = 5$ kHz is bended and absorbed by the proposed device, and the patterns are almost the same as the ideal continuous model with $\delta = 0.15 \times (\frac{R'}{r'})^2$ in the shell region, which is shown in Fig.3(c). Here we have to note that in this scheme, the bandwidth is limited by the cell structure, because the quasi-static condition have to be guaranteed to make the effective material parameters accurate, this requires the wavelength is large enough compared to the scale of the cell element. To enlarge the bandwidth, a model with relatively more layers is required, which we think is not a major problem for the manufacturing technique of today. In addition, for such omnidirectional broadband elastic wave absorber, there is sufficient flexibility in the design process. For example, with other transformation relations reported in [18], similar device constituted by only two kinds of materials can also be expected.

In conclusion, based on a conformal mapping and transformation method, we have proposed a general approach to design "black-hole" absorbers. A series of omnidirectional absorbers are proposed for different physical waves, especially for elastic wave absorber. A prototype of an elastic wave omnidirectional absorber is also proposed and verified by numerical simulation, the possible applications in wave absorption and energy harvesting can be anticipated.

Acknowledgments:
This work was supported by the National Natural Science Foundation of China (10832002), and the National Basic Research Program of China (2006CB601204).

Figures:

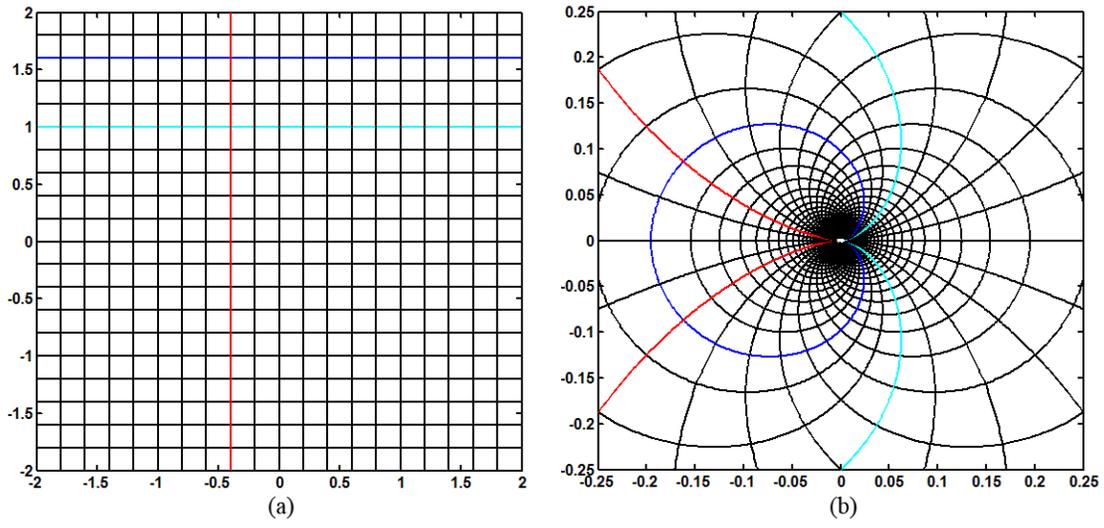

Figure 1. (Color Online) Schematic diagram of a flat virtual space $z = x + iy$ (a) and a curved physical space $w = x' + iy'$ (b) transformed by the mapping $w = \frac{1}{2z^2}$.

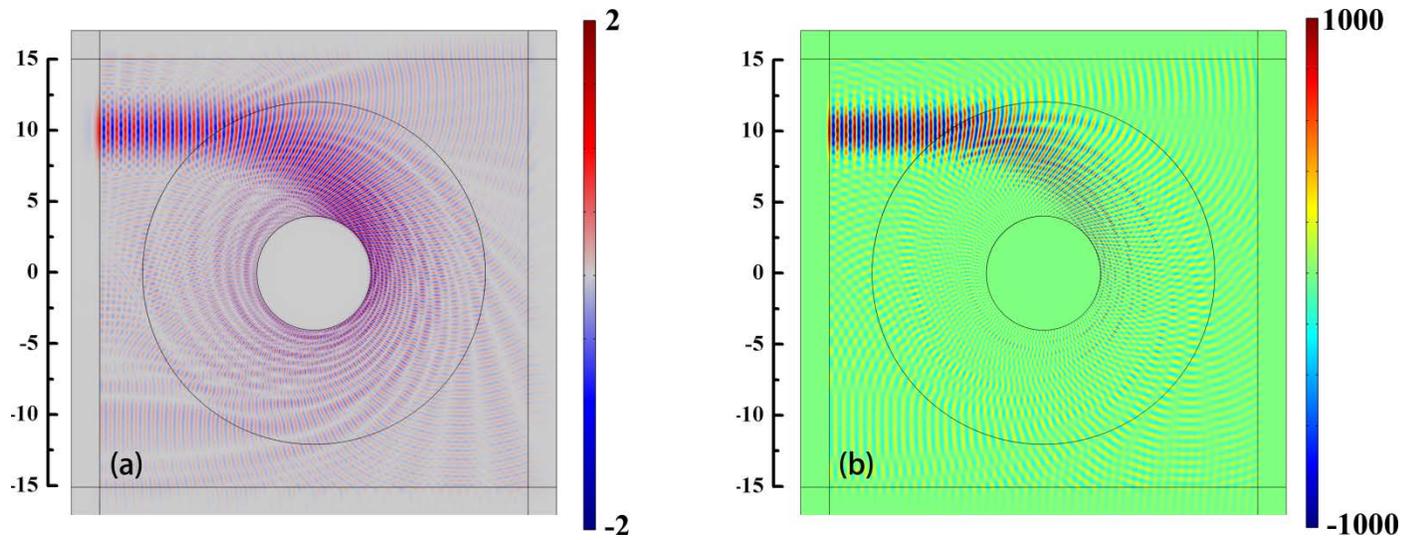

Figure 2. (Color Online) (a) The electric field when a Gaussian beam ($f = 100\text{GHz}$) is incident on the magnetic "optical black-hole". (b) The sound pressure field when a Gaussian sound beam ($f = 2.5\text{MHz}$) is incident on the mass-density-controlled "acoustic black-hole". They show similar patterns although the fields are different.

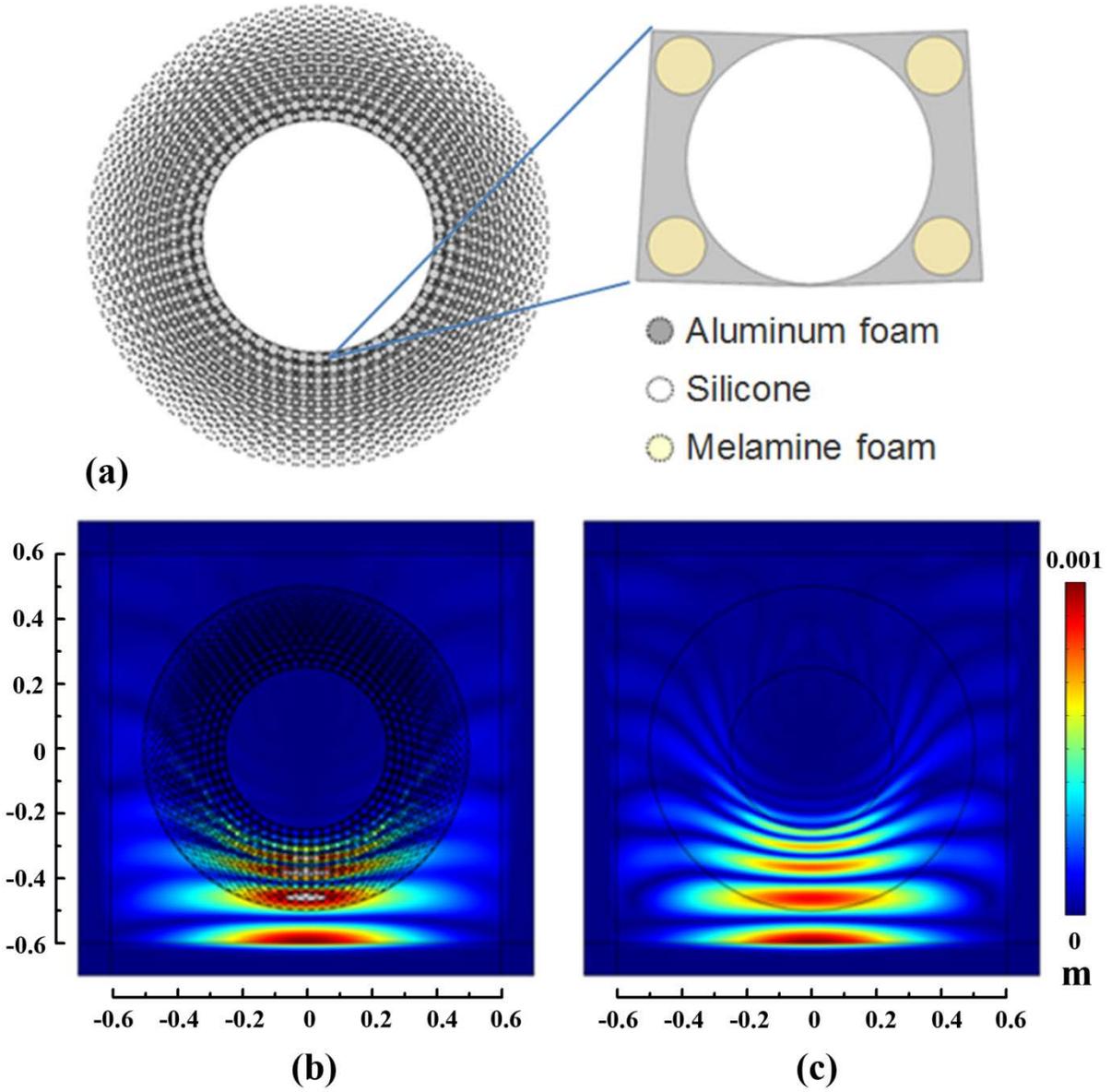

Figure 3. (Color Online) (a) the implementation scheme of an "elastic wave black-hole". The "black-hole" region is spliced by the cell elements consisting of aluminum foam, silicone and melamine foam. The total displacement field of the designed "black-hole" device (b) shows the incident S-wave beam was bended and absorbed in the device, which has a similar result with that of the continuous model (c).